\documentclass{svproc}
\usepackage{svg}
\usepackage{graphicx}%
\usepackage{multirow}%
\usepackage{amsmath,amssymb,amsfonts}%
\usepackage{mathrsfs}%
\usepackage[title]{appendix}%
\usepackage{xcolor}%
\usepackage{textcomp}%
\usepackage{manyfoot}%
\usepackage{booktabs}%
\usepackage{listings}%
\usepackage{url}

\def\orcidID#1{\unskip$^{[#1]}$} 

\usepackage{subcaption}
\usepackage{cleveref}
\usepackage{todonotes} 
\usepackage{multirow}
\usepackage[most]{tcolorbox}
\usepackage[linesnumbered,ruled,vlined]{algorithm2e}

\begin{document}
\mainmatter              
\title{DynBenchmark: Customizable Ground Truths to Benchmark Community Detection and Tracking in Temporal Networks} 
\titlerunning{DynBenchmark}  
%
\author{Laurent Brisson\inst{1}\orcidID{0000-0002-5309-2688}, Cécile Bothorel\inst{1}
\and Nicolas Duminy\inst{1}}
\authorrunning{Laurent Brisson et al.} 
%
\institute{IMT Atlantique, Lab-STICC, UMR CNRS 6285, Brest, France\\
\email{laurent.brisson@imt-atlantique.fr}
}

\maketitle              

\begin{abstract}
Graph models help understand network dynamics and evolution. Creating graphs with controlled topology and embedded partitions is a common strategy for evaluating community detection algorithms. However, existing benchmarks often overlook the need to track the evolution of communities in real-world networks. To address this, a new community-centered model is proposed to generate customizable evolving community structures where communities can grow, shrink, merge, split, appear or disappear. This benchmark also generates the underlying temporal network, where nodes can appear, disappear, or move between communities. The benchmark has been used to test three methods, measuring their performance in tracking nodes' cluster membership and detecting community evolution. Python libraries, drawing utilities, and validation metrics are provided to compare ground truth with algorithm results for detecting dynamic communities.
\keywords{Community Detection,
Temporal Networks,
Evolving Communities,
Ground Truth,
Planted Communities,
Benchmark}
\end{abstract}
\section{Introduction}
Community discovery is crucial in analyzing complex networks, in order to identify mesoscale structures that describe network organization. In temporal networks, which represent, for example, interactions over time, nodes and edges appear or disappear over time, affecting community topology. During the past decade, time-aware methods have been developed to detect evolving community structures \cite{rossetti2018community,DAKICHE20191084,Christopoulos2022}. However, validating these methods remains challenging due to the lack of ground truth for real datasets. Although some academic repositories provide metadata as ground truth \cite{clauset:2016,rossi:2015a,jerome:2013a,Leskovec:2019}, this information often describes node attributes rather than network structure, making it an unreliable benchmark for evaluating community detection algorithms \cite{yang:2012a,dao:2017b,peel:2017a}. 

Graph generators that reproduce real-world linking properties are another way to measure the effectiveness of graph algorithms. Bonifati et al. provides an overview of the state-of-the-art graph generators \cite{Bonifati:2020} but unfortunately, even though the area of dynamic graphs is intensively studied, surprisingly there seem to exist only very few proposals of a generator for dealing with dynamics, and especially with community dynamics.

Some approaches take an initial partition generated by a static algorithm and let it evolve randomly. They generally adapt classical benchmarks for static networks to dynamic networks: Lin et al. \cite{Lin:2009:ACE:1514888.1514891} build on Girvan and Newman's one \cite{girvan2002community} where nodes have fixed degree and communities have fixed size; Folino and Pizzuti \cite{folino2014evolutionary} propose an adaptation of the LFR benchmark originally defined by Lancichinetti et al.  \cite{lancichinetti2008benchmark} with a power-law degree distribution and different community sizes. 
On the contrary, for their initial partition, Granell et al. don't use an existing static generator but start with $q$ communities with $n$ nodes equally distributed among the communities. The dynamics is a periodic oscillation of communities changes, with combinations of growing/shrinking and/or merging/splitting processes \cite{granell2015benchmark}. The main limitation of these benchmarks is a constant number of nodes. This strong constraint leads to rigid operations on communities. Communities are all of the same size, and if not, their size varies in the same proportions. Evolutionary behaviors, when they are considered, are simulated in a regular and periodic way. Operations such as birth or death are not taken into account, or, if they are, new communities are created from the absorption of existing nodes.

RDyn \cite{Rossetti2017} offers more diversity, with community size and node degree following a power-law distribution (power-law distribution), but again the number of nodes is constant. In addition, the model is designed to identify patterns of evolution (e.g., merging, splitting), but not to track the evolution of communities over time, i.e. to build a sequence of static communities, one per snapshot, as an evolving community. The validation of algorithms is done snapshot per snapshot without any metrics to evaluate the dynamics. 
A recent benchmark called Mosaic \cite{asgari2023mosaicbenchmarknetworksmodular} takes a different approach by generating ground truth using the link streams formalism. While this approach effectively simulates sparse interaction patterns and can evaluate algorithms' ability to reconstruct communities from high-resolution temporal data, it focuses primarily on community detection accuracy rather than evolution dynamics.


The authors argue that benchmark models for community detection in temporal networks need to be more realistic. They propose a new benchmark model focused on community evolution, with the following key contributions:
\begin{enumerate}
    \item \textbf{Diversity}: The model generates diverse evolutionary communities with varying sizes and lifetimes, including the appearance and disappearance of nodes. It also creates artificial networks that match these communities.
    \item \textbf{Customizability}: The model is highly configurable, with many input parameters controlling community dynamics and network evolution. These parameters are defined by probability distributions, allowing for stochastic and dynamic ground truth communities.
    \item \textbf{Availability}: The authors provide Python libraries, drawing utilities, and validation metrics to compare ground truth with the results of dynamic community detection algorithms: partitions, transitions, events \cite{gitlab}.
\end{enumerate}

The rest of the article is organized as follows. First, we present in section 2 our model and the measures to compare dynamic community detection algorithms to ground truths. In Section 3 we assess the ground truth itself, and in Section 4 we conduct some experiments to show the application of the proposed benchmark. Finally, we give a summary and some perspectives. 

\section{Benchmark description}

Let us consider a temporal network that operates over a duration encompassing \(T\) discrete times. We define a snapshot $G_t$ as an undirected and unweighted static graph at time $t$, representing the active nodes and interactions at that time. Each snapshot $G_t$ can be partitioned into multiple communities. We denote by $C_{k,t}$ a specific static community $k$ at time step $t$. Within this framework, an evolving community, denoted as \(C_k\), persists over multiple consecutive time steps. \(C_k\) is defined as a sequence of static communities $C_{k,t}$.


\subsection{Ground truth generator}

As our benchmark is a community-centred model, the first step is to generate empty evolving groups and their interactions before assigning nodes and edges:

\begin{enumerate}
    \item \textbf{Generation of evolving communities:} Users provide probability distributions to regulate the number of evolving communities, their lifespan, and the distribution of their birth within a specified time window. Additionally, the evolution of these communities over time can be precisely controlled through user-defined distributions: changes in size and member turnover (the core nodes ratio defining the flow between \(C_{k,t}\) to community \(C_{k,t+1}\), 
    
    \item \textbf{Assignment of members to static communities \(C_{k,t}\):} This step involves assigning members to each static community, according to the flows fixed in the previous step. 
    
    \item \textbf{Generation of underlying graphs $G_t$:} The final step revolves around generating the underlying graphs that depict the relationships between members for each snapshot. We use the Stochastic Block Model (SBM) with controllable intra-community and inter-community link densities provided by the users.
\end{enumerate}


This highly configurable approach enables researchers to generate diverse temporal network scenarios. Researchers can systematically vary parameters to examine algorithm performance under different underlying graph topologies (link densities), community evolution dynamics (member flows, persistence), and structural properties (community size, lifespan, quantity). This capability is particularly valuable for identifying algorithm strengths and weaknesses across diverse temporal network characteristics.

\subsection{Algorithms Assessment}

Our benchmark framework enables rigorous evaluation of dynamic community detection algorithms along three complementary dimensions:  (1) the quality of the detected partitions at each timestep, (2) their ability to accurately track community transitions between snapshots, and (3) their capacity to identify key evolutionary events in the communities' life-cycle. In this section, we focus particularly on the latter two dimensions that specifically address temporal dynamics, demonstrating how our framework reveals strengths and limitations of algorithms in tracking evolving community structures.

\paragraph{Nodes Transitions}


Node transitions between communities enable the analysis of community dynamics over time. As proposed by \cite{granell2015benchmark}, let $\mathcal{P}_t$ and $\mathcal{P}_{t+\delta}$ be the community partitions of two snapshots $G_t$ and $G_{t+\delta}$ of the graph, where $\delta$ is a strictly positive time offset. For each active node $v$ in both snapshots, we define its transition as a pair in $T_{t,\delta} = \mathcal{P}_t \times \mathcal{P}_{t+\delta}$, where the pair elements are respectively the communities of node $v$ at times $t$ and $t+\delta$. This representation allows building a contingency matrix $M$ comparing observed transitions with ground truth ones, where each element $m_{i,j}$ corresponds to the number of nodes sharing the same transitions. To compare such transitions, various similarity measures like Normalized Mutual Information (NMI) and Normalized Variation of Information (NVI) are commonly used. To better handle the dynamic nature of temporal networks, \cite{granell2015benchmark} propose windowed versions of these measures that compare sequences of partitions by computing the mean squared error over a specified time window.

\paragraph{Communities events}

An essential aspect of analyzing community dynamics involves tracking the events that shape the life cycle of communities. The evaluation of dynamic community detection algorithms requires understanding how these communities evolve through specific events (e.g., form, continue, merge and grow, partial merge, dissolve). As part of our benchmark, we incorporate an implementation of  ICEM algorithm proposed by Mohammadmosaferi et al. \cite{KadkhodaMohammadmosaferi2020}.

\section{Resulting ground truths}\label{sec:ground_truth}

According to Table \ref{tab:exp-params}, we generated ground truths where 10 evolving communities run on 10 snapshots, with a minimum size of 10 members. The size of the communities at birth is governed by a normal (Gaussian) distribution, with a mean of 50 and a standard deviation of 20.
Their lifetime is governed by a truncated normal distribution, with a mean of 5 and a standard deviation of 2. According to their picked lifetime, they randomly start at $t \in [0,9- PickedLifetime]$.  These base parameters remained constant across all scenarios, while we systematically explored different combinations of graph and community parameters, resulting in 96 unique configurations. Each configuration was tested with 100 different instances, resulting in 9,600 ground truths. 

\begin{table}[h!]
\centering
\caption{Experimental Parameters (100 instances per configuration)}
\label{tab:exp-params}
\begin{tabular}{lll}
\toprule
\multirow{6}{*}{\textbf{\begin{tabular}[c]{@{}l@{}}Base\\configuration\end{tabular}}} 
& Number of timesteps & 10 \\
& Number of communities & 10 \\
& Minimum community size & 10 \\
& Initial community size & $\mathcal{N}(50, 20)$ \\
& Community start time & $\mathcal{U}[0,1]$ \\
& Community lifetime & $\mathcal{N}_{[3,7]}(5, 2)$ \\
\midrule
\multirow{4}{*}{\begin{tabular}[c]{@{}l@{}}\textbf{Parameter Space}\\(96 unique configurations)\end{tabular}}
& $p_{in}$ & \{0.25, 0.5, 0.75\} \\
& $p_{out}$ & \{0.025, 0.05, 0.075, 0.1\} \\
& Size change ratio & \{0, $\mathcal{N}(0, 0.2)$\} \\
& Core nodes ratio & \{0.25, 0.5, 0.75, 1.0\} \\
\bottomrule
\end{tabular}
\end{table}

\begin{figure}[h!]
    \centering
    \begin{subfigure}{0.49\linewidth}
        \centering
        \includegraphics[width=0.95\linewidth]{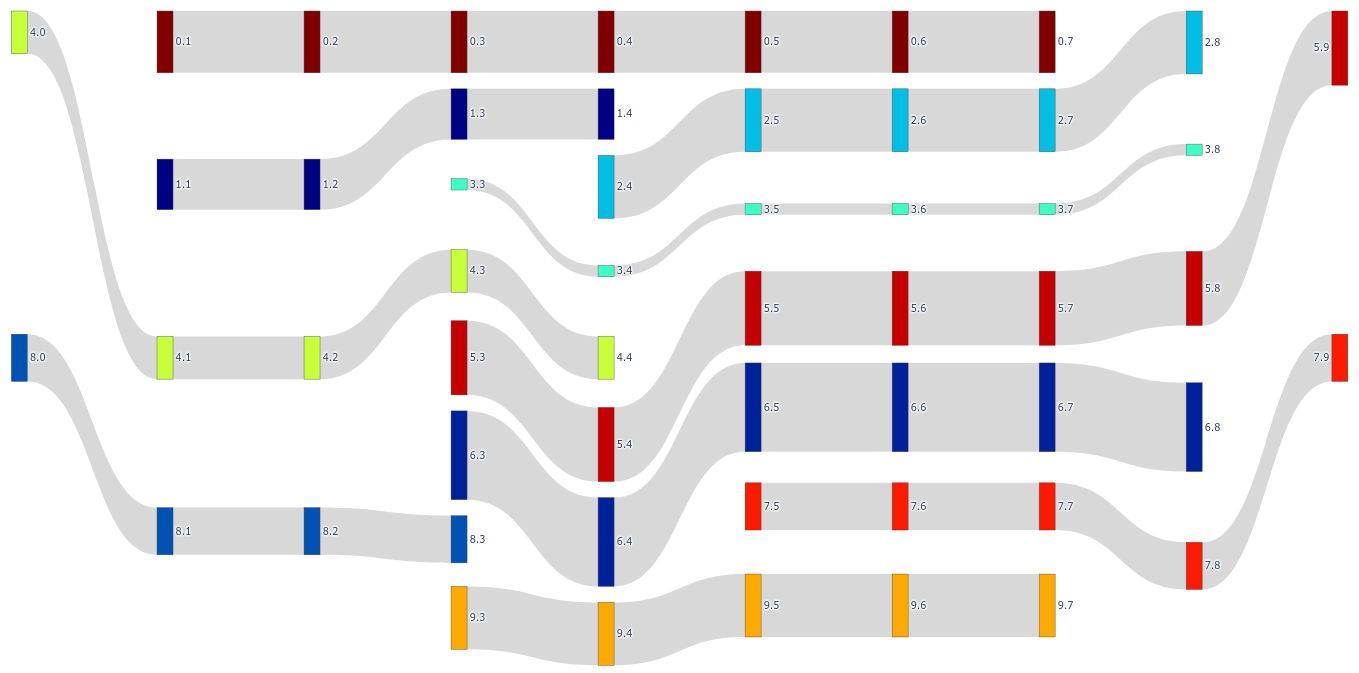}
        \caption{Baseline\_1.0}
        \label{fig:baseline_1.0-sankey}
    \end{subfigure}
    \begin{subfigure}{0.49\linewidth}
        \includegraphics[width=0.95\linewidth]{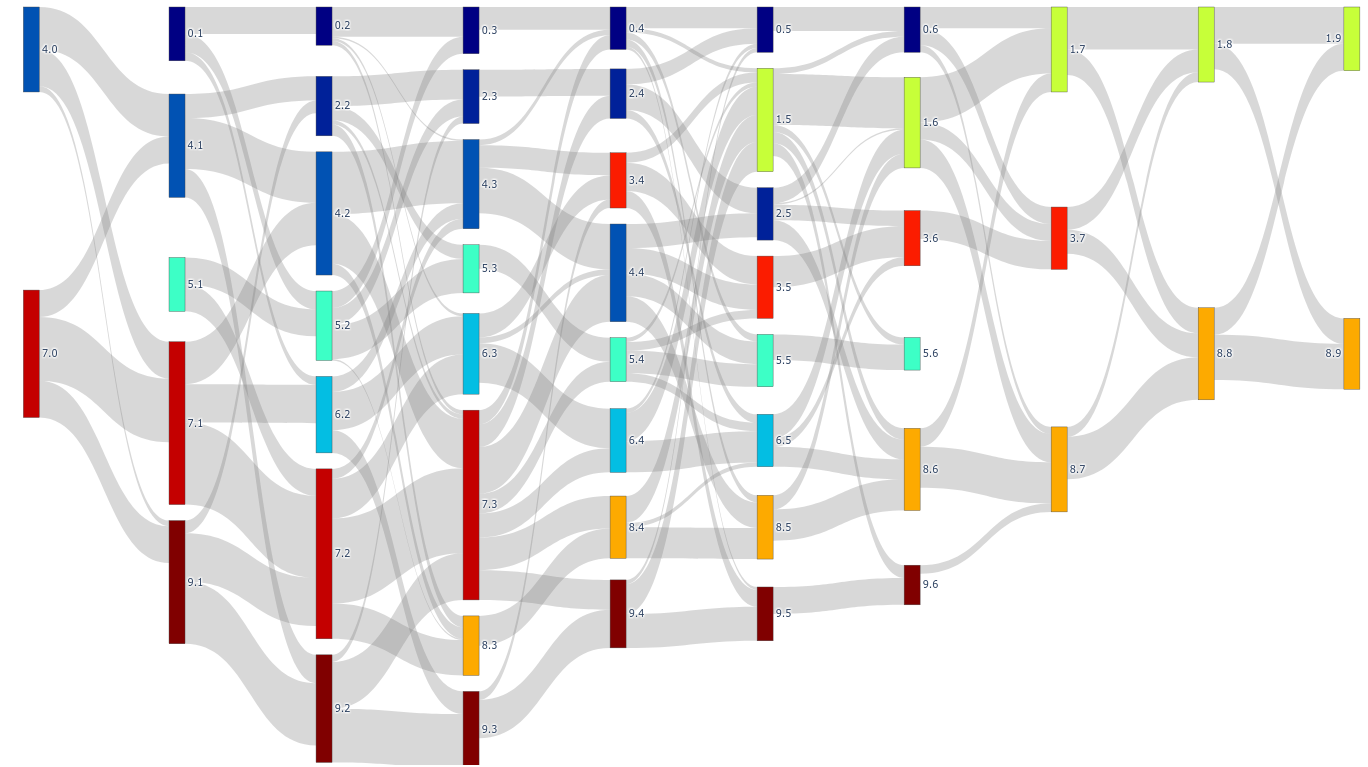}
        \label{fig:changing_0.5-sankey}
        \caption{Changing\_0.5}
    \end{subfigure}%
    \caption{Communities' lifespan and interactions. Colors differentiate communities, columns reflect time (10 snapshots). The thickness or thinness of the gray flows is indicative of the number of members migrating between communities. }
    \label{fig:sankeys}
\end{figure}

\paragraph{Evolving communities} The 9,600 experiments generated a total of 96,000 communities (10 per experiment as expected). 
They evolve in parallel during an average of 5.67 snapshots, but with a wide variety: the snapshots contain 1 to 10 communities, with more than 6 of them in a majority of time steps. A total of 529,248 static communities are created, with 50 members on average (standard deviation 23.37, minimum and maximum sizes are 10 and 217 respectively). 
As shown in Fig. \ref{fig:sankeys}, using distributions introduces diversity in lifetime, initial sizes, and time of birth. 
To test the variety of behaviors, we defined 2 scenarios: 
\begin{itemize}
    \item \textit{Changing\_cnr}: We use a normal distribution to apply a size change ratio ($\mu=0.0$, $\sigma=0.2$). To control the stability of a community in terms of members, we define a core node ratio $cnr \in [0.25, 0.5, 0.75, 1.0]$  that guarantees that a percentage of members remain in the community from one timestamp to the next.
    \item \textit{Baseline\_cnr}: The size change ratio is fixed at zero to get stable communities. We set the same 4 core node ratios. With a core node ratio  $cnr$ equal to 100\%, we obtain very stable communities, with no member turnover as depicted in Fig. \ref{fig:baseline_1.0-sankey}.
\end{itemize}

Table \ref{tab:groundtruth:dynamics} details the dynamics of community evolution. Member turnover aligns with the core node ratio, showing that members stay in their communities according to the defined ratio. The most stable scenario, \textit{baseline\_1.0}, has no emigration, and each community remains its own predecessor. In contrast, a high-turnover community like \textit{baseline\_0.25} sees significant member movement, with 75\% of new members coming from multiple predecessors (itself and 3.32 others). When communities are born or die, members are reused or disappear, affecting platform turnover measured by the System renewal indicator. As expected, minor size fluctuations in \textit{changing} size scenarios have minimal impact on emigrant ratio, turnover, and the number of predecessors.

\begin{figure*}[h!]
    \centering
    \begin{subfigure}{0.49\linewidth}
        \includegraphics[width=\linewidth]{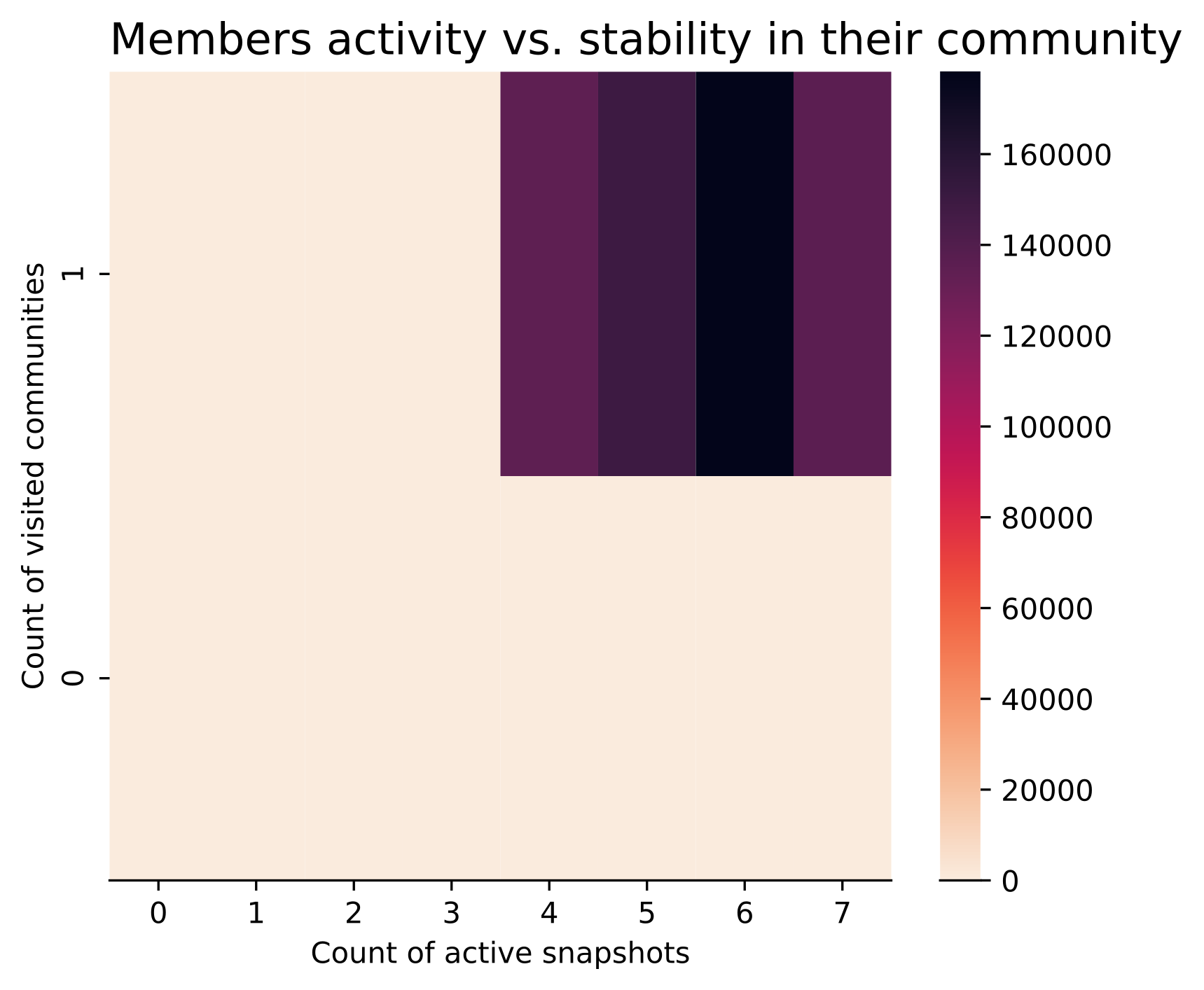}
        \caption{Baseline\_1.0}
        \label{members:baseline-1.0}
    \end{subfigure}
    \begin{subfigure}{0.49\linewidth}
        \includegraphics[width=\linewidth]{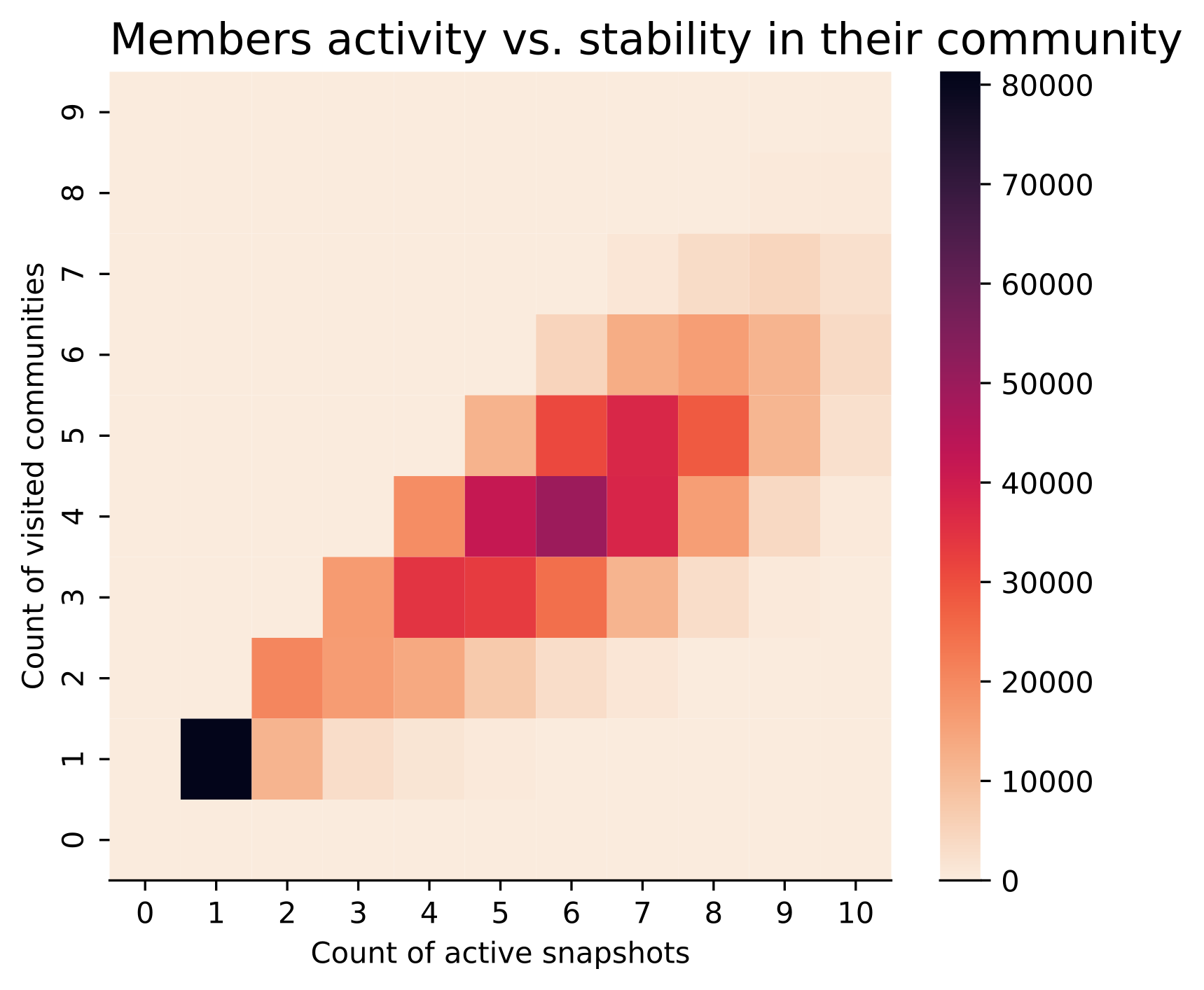}
        \caption{Baseline\_0.25}
        \label{members:baseline-0.25}
    \end{subfigure}
    \caption{Impact of core node ratio on members' trajectories: number of communities visited by each node and their lifetime}
    \label{default_degree_membership_turnover}
\end{figure*}

The core node ratio influences member trajectories within communities. When the core ratio is at its maximum (100\%), nodes remain in the same community and stay active throughout its lifespan (Fig. \ref{members:baseline-1.0}). Conversely, a lower core ratio (e.g., 0.25\%) results in unstable members who move between multiple communities, have shorter lifespans, and may disappear from the dataset more quickly  (Fig. \ref{members:baseline-0.25}). 

\begin{table*}[h!]
\centering
\caption{Community dynamics}
\label{tab:groundtruth:dynamics}
\begin{tabular*}{0.96\textwidth}{@{\extracolsep{\fill}} lrrrrrrrr}
\toprule
{} & Size   & Emigrant  & Turnover  & System   & Nb of  \\
{} &  change &  ratio &  ratio & renewal  &  predecessors\\
\midrule
baseline\_0.25 &  $0.00\pm0.00$   & $0.58\pm0.28$   &  $0.75\pm0.10$   & $0.27\pm0.35$     &  $4.32\pm1.47$ \\
baseline\_0.50 &  $0.00\pm0.00$   & $0.39\pm0.27$   &  $0.50\pm0.01$   & $0.22\pm0.31$     &  $2.65\pm0.94$\\
baseline\_0.75 &  $0.00\pm0.00$   & $0.20\pm0.23$   &  $0.25\pm0.01$   & $0.21\pm0.31$     &  $1.98\pm0.83$ \\
baseline\_1.0 &   $0.00\pm0.00$   & $0.00\pm0.00$   &  $0.00\pm0.00$   & $0.22\pm0.31$     &  $1.00\pm0.00$ \\
\midrule
changing\_0.25   &  $0.00\pm0.20$   & $0.57\pm0.29$   &  $0.75\pm0.10$  & $0.27\pm0.36$ &  $4.15\pm1.61$ \\
changing\_0.50   &  $0.00\pm0.20$   & $0.38\pm0.27$   &  $0.50\pm0.10$  & $0.22\pm0.33$ &  $2.57\pm1.06$ \\
changing\_0.75   &  $0.00\pm0.20$   & $0.20\pm0.23$   &  $0.26\pm0.08$  & $0.21\pm0.33$ & $1.90\pm0.88$ \\
changing\_1.0    &  $0.00\pm0.20$   & $0.07\pm0.16$   &  $0.08\pm0.06$  & $0.22\pm0.33$ &  $1.37\pm0.67$ \\
\bottomrule
\end{tabular*}
\end{table*}

\begin{figure*}[h!]
\centering
    \centering
    \begin{subfigure}{0.45\linewidth}
        \includegraphics[width=\linewidth]{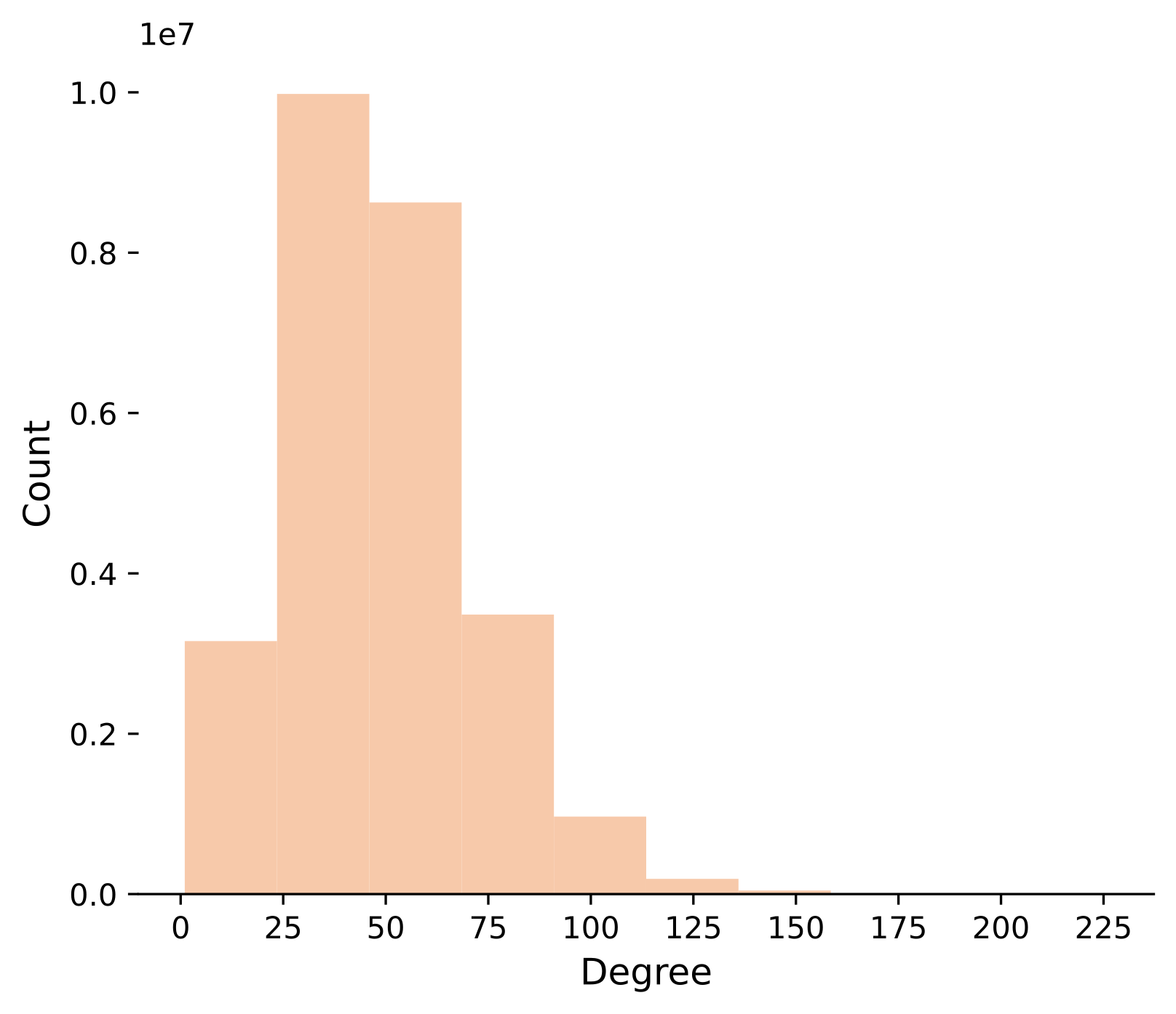}
        \caption{Degree distribution}
        \label{fig:degree}
    \end{subfigure}
    \begin{subfigure}{0.45\linewidth}
        \includegraphics[width=\linewidth]{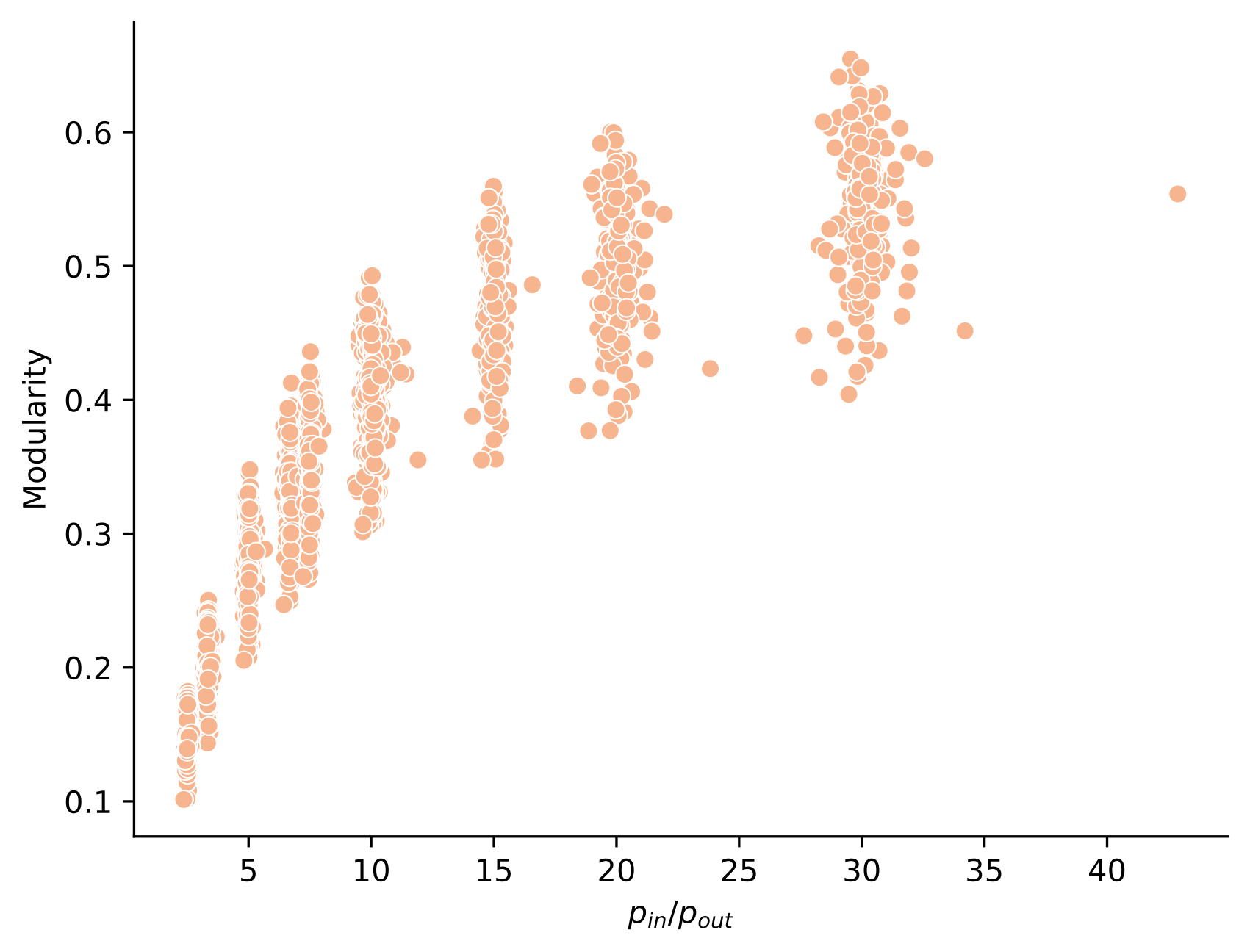}
        \caption{Modularity vs. $p_{in}/p_{out}$}
         \label{fig:mixing_parameter}
    \end{subfigure}
    \caption{Properties of the underlying networks}
\end{figure*}

\paragraph{Underlying networks} We generate the 93,360 underlying networks with the SBM model to get clustered nodes. We control communities' internal and external edges densities, but not the distribution of edges on nodes (like the preferential attachment, for example). So, as expected, the degree distribution doesn't follow a power law (Fig. \ref{fig:degree}), thus failing to satisfy one of the properties of scale-free networks. But the average shortest paths are very low (mean $1.88\pm0.19$) leading to low diameters (from 2 to 9, mean $3.03\pm0.42$) and then fitting small-world network properties.

\begin{table*}[h!]
\centering
\caption{Properties of the 93,360 generated networks}
\label{tab:networks}
\begin{tabular*}{0.95\textwidth}{@{\extracolsep{\fill}} lrrrrrrrr}
\toprule
{} &  diameter &  \# of nodes &  \# of edges &  average shortest path &      ccf  \\
\midrule
mean  &      3.03 &         283.44 &   6873.98 &   1.88 &   0.31  \\
std   &      0.42 &          145.24 &  5599.18 &   0.19 &     0.17 \\
min   &      2.00 &          10.00 &   12.00 &      1.20 &     0.00  \\
max   &      9.00 &          721.00 &   50182.00 &   3.17 &     0.80 \\
\bottomrule
\end{tabular*}
\end{table*}

We generated 12 scenarios with different levels of community structure difficulty (see Table \ref{tab:exp-params}). With these different combinations of $p_{in}$ and $p_{out}$ we control the fraction of internal and external edges. According to the Fig. \ref{fig:mixing_parameter}, networks exhibit clustered partitions when the resulting densities respect $p_{in}>>p_{out}$. Nodes primarily connect to others within their community, resulting in well-defined and easily detectable communities.

\section{Assessing temporal clustering methods}

To demonstrate our benchmark's capability to evaluate community detection algorithms through time, we performed an evaluation across multiple experimental configurations (cf. Table \ref{tab:exp-params}).
We analyzed three community detection algorithms---Louvain, Infomap and Walktrap---adapted to temporal networks through community matching techniques. Each algorithm was assessed against our 9,600 ground truths. Our evaluation examines three dimensions: (1) partition quality, (2) transition tracking accuracy, and (3) evolutionary event detection. Our package implements similarity metrics to compare detected sets with their corresponding ground truth: the Adjusted Rand Index (ARI), F1-score, Normalized Mutual Information (NMI), Normalized Variation of Information (NVI), and Jaccard index.

\subsection{Assessing Partition Quality} 

Our benchmark enables sensitivity analysis to generation parameters. Results presented here use the \textit{changing\_0.5} scenario averaged across all experimental runs.

\begin{figure}[ht!]
\centering
\begin{subfigure}[b]{0.45\textwidth}
    \centering
    \includegraphics[width=\textwidth]{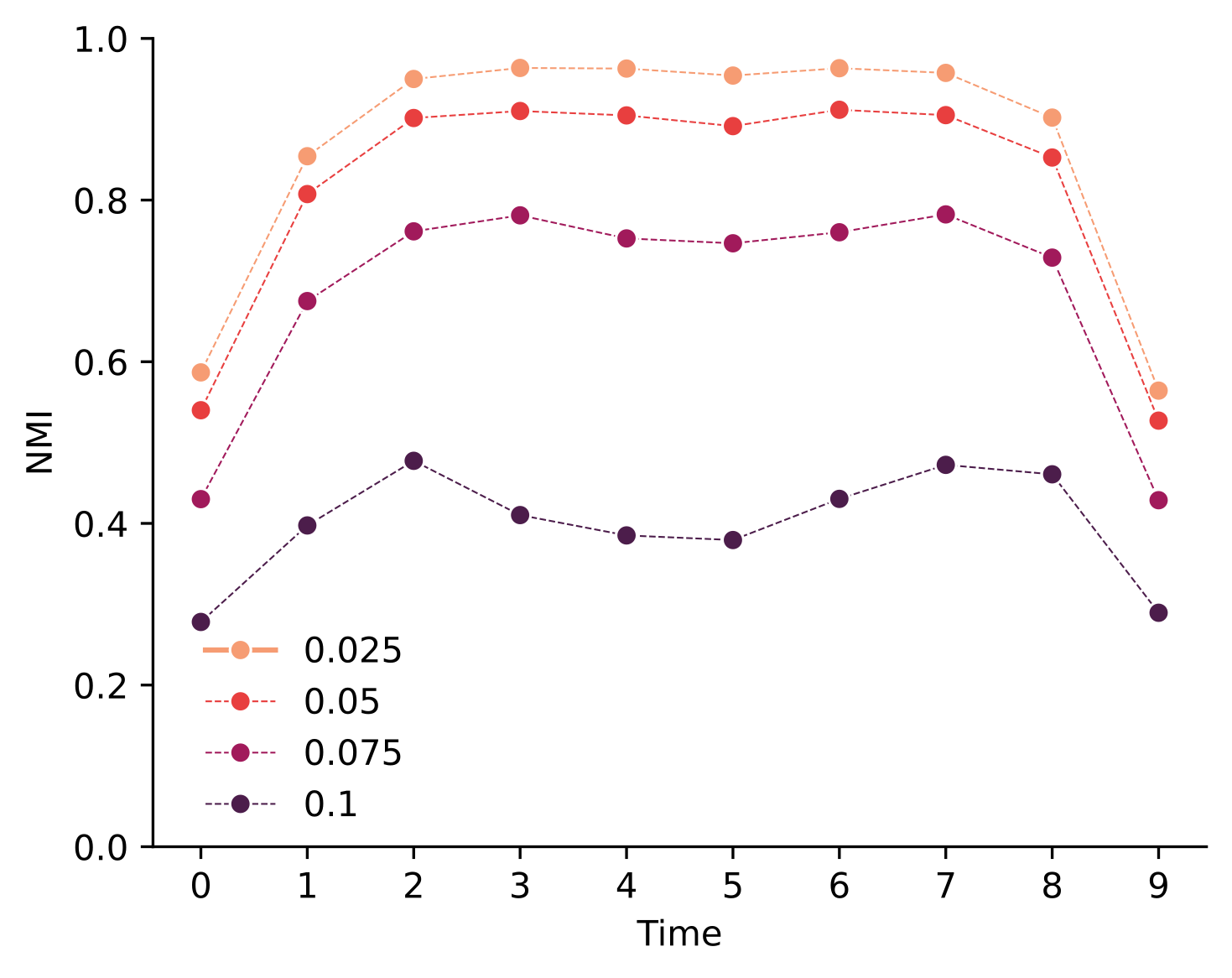}
    \caption{Impact of $p_{out}$ on Louvain}
    \label{fig:louvain-overtime}
\end{subfigure}
\hfill
\begin{subfigure}[b]{0.45\textwidth}
    \centering
    \includegraphics[width=\textwidth]{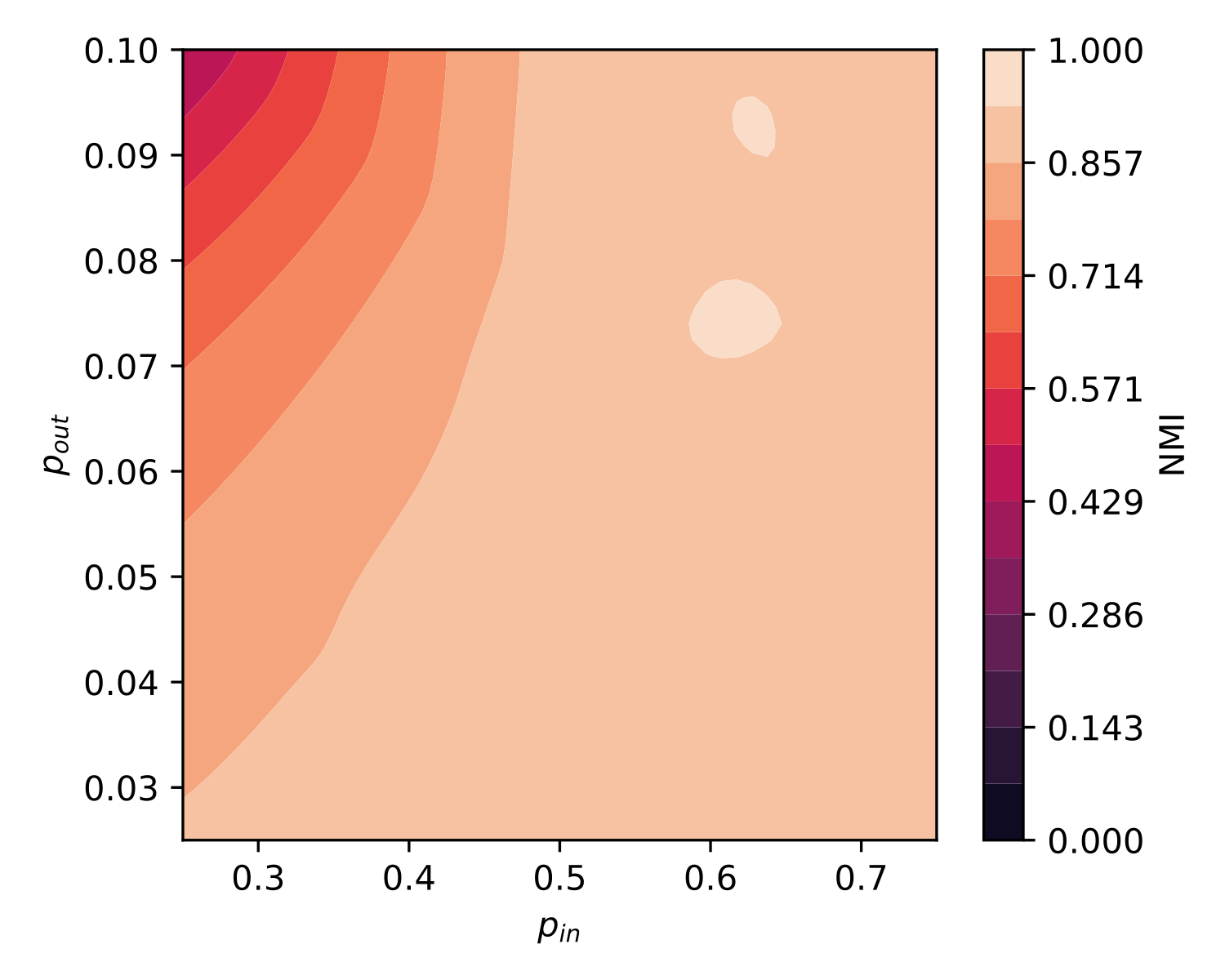}
    \caption{Louvain }
    \label{fig:louvain-heatmap}
\end{subfigure}
\begin{subfigure}[b]{0.45\textwidth}
    \centering
    \includegraphics[width=\textwidth]{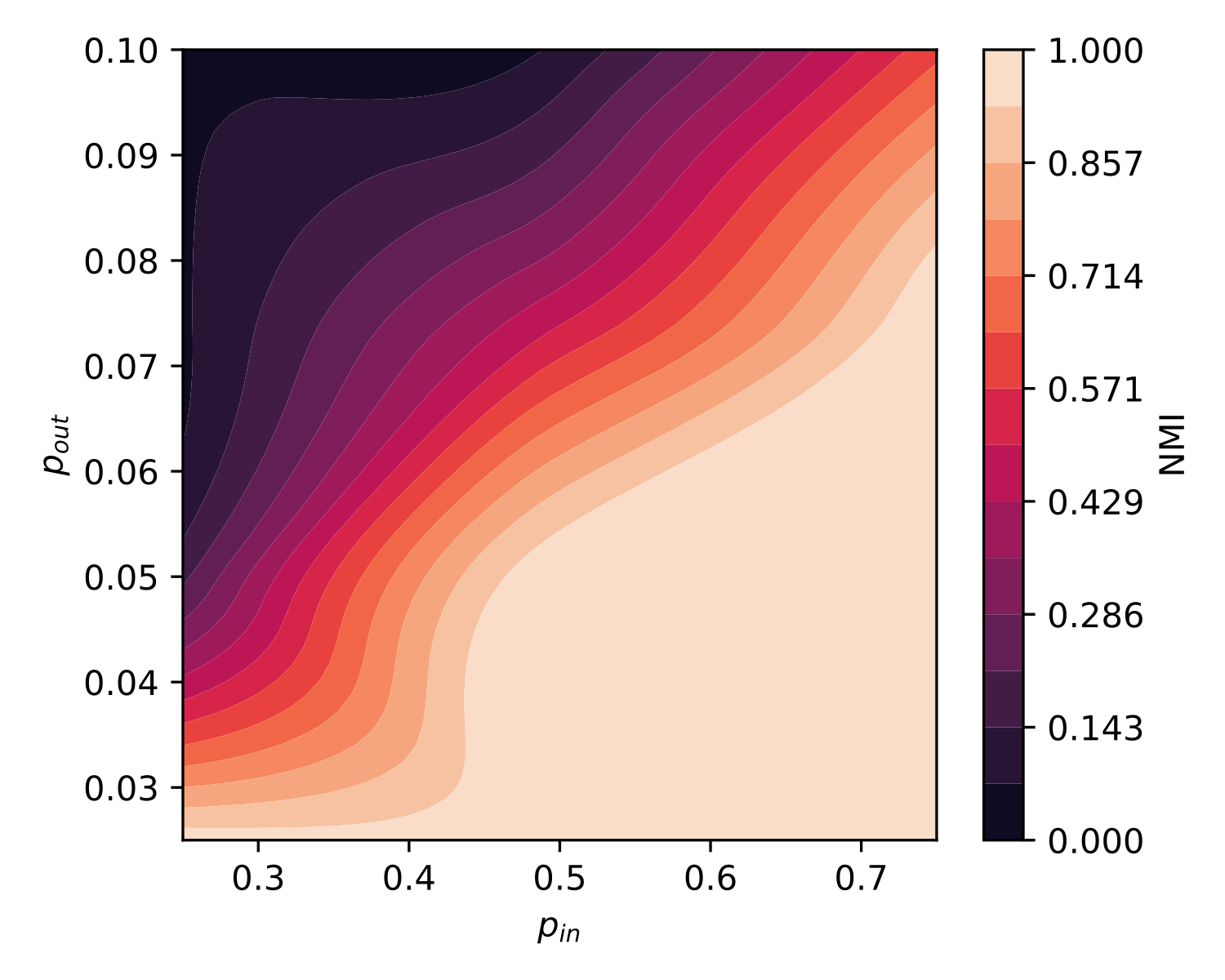}
    \caption{Infomap }
    \label{fig:infomap-heatmap}
\end{subfigure}
\hfill
\begin{subfigure}[b]{0.45\textwidth}
    \centering
    \includegraphics[width=\textwidth]{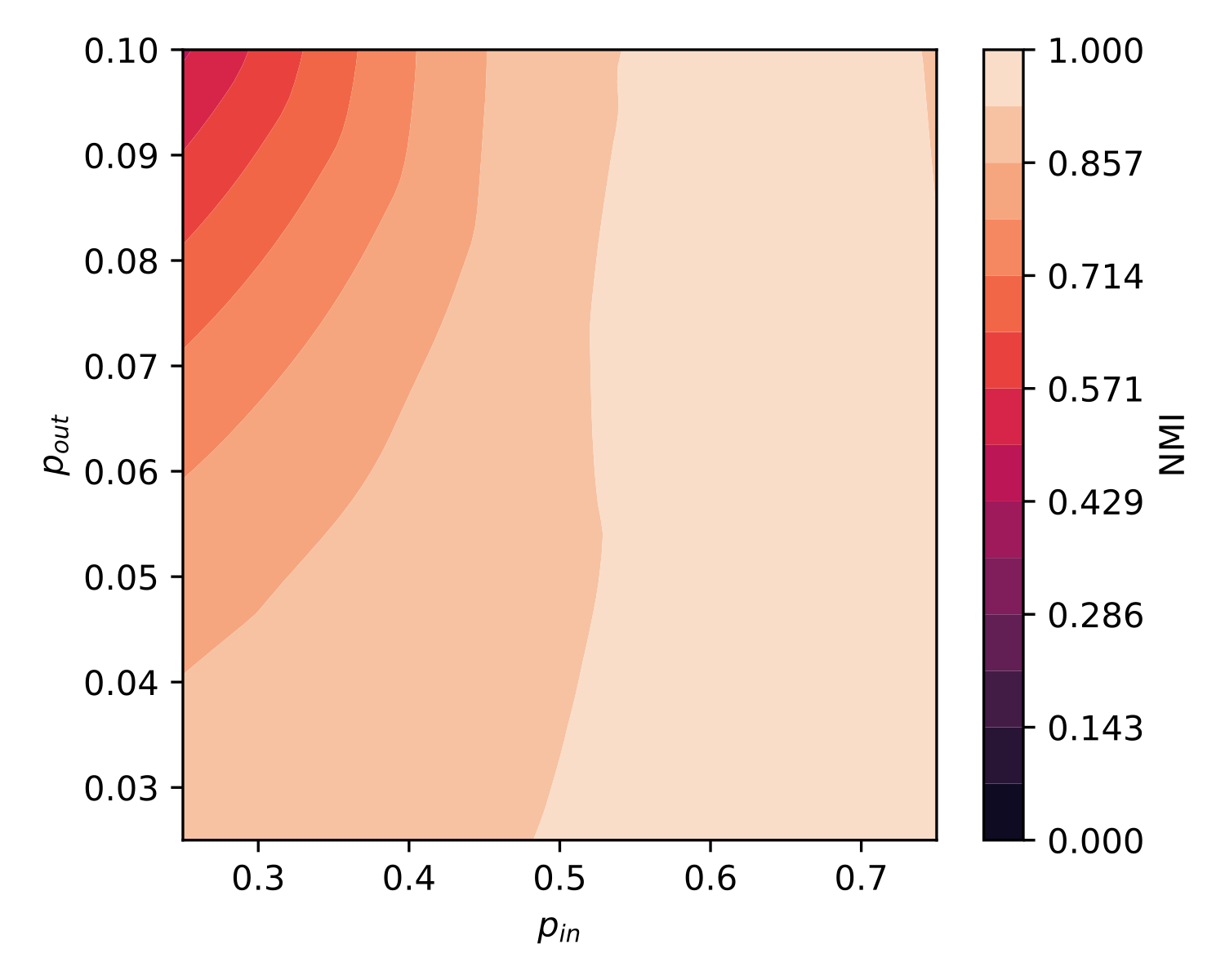}
    \caption{Walktrap}
    \label{fig:walktrap-heatmap}
\end{subfigure}
\caption{(a) Evolution of Louvain's performance (NMI) over time with fixed internal connectivity ($p_{in}=0.5$) and different values of $p_{out}$. (b-d) Sensitivity analysis showing the impact of both $p_{in}$ and $p_{out}$ on the performance of Louvain, Infomap, and Walktrap respectively.} 
\label{fig:partitions-comparison}
\end{figure}

As an example, Fig. \ref{fig:partitions-comparison}\subref{fig:louvain-overtime} illustrates how Louvain performs with a fixed internal connectivity $p_{in}=0.5$ and varying external connection probabilities ($p_{out}$). The algorithm maintains good performance ($NMI > 0.8$) when inter-community connections remain sparse ($p_{out} \leq 0.05$). As expected, performance degrades when $p_{out}$ increases, reflecting communities that are too interconnected and therefore indistinguishable.

Fig. \ref{fig:partitions-comparison}\subref{fig:louvain-heatmap}, \ref{fig:partitions-comparison}\subref{fig:infomap-heatmap} and \ref{fig:partitions-comparison}\subref{fig:walktrap-heatmap} examine how  ($p_{in}$) and ($p_{out}$) affect the performance of each algorithm. The analysis shows distinct performance patterns for three community detection algorithms: Infomap is highly sensitive to parameters, with performance dropping as external connectivity $p_{out}$ increases. It requires high internal ($p_{in} > 0.5$) and low external ($p_{out} < 0.06$) connectivity to maintain accuracy. Louvain is more robust, with a gradual performance decline as the network becomes more randomized.  Walktrap is the most resilient, maintaining high accuracy and perfect concordance with the ground truth when internal connectivity is greater than 0.5, regardless of external connectivity.


\subsection{Assessing Transition Accuracy} 

Beyond the classical partition quality assessment, our benchmark evaluates how well algorithms capture the evolution of communities over time. By analyzing collective node movements between snapshots, we can quantify the algorithms' ability to track membership changes. Focusing here the scenario \textit{changing\_0.5}, we examine how temporal distance and algorithm choice impact community tracking performance.

\begin{figure}[h]
\centering
\begin{subfigure}[b]{0.48\textwidth}
\centering
\includegraphics[width=\textwidth]{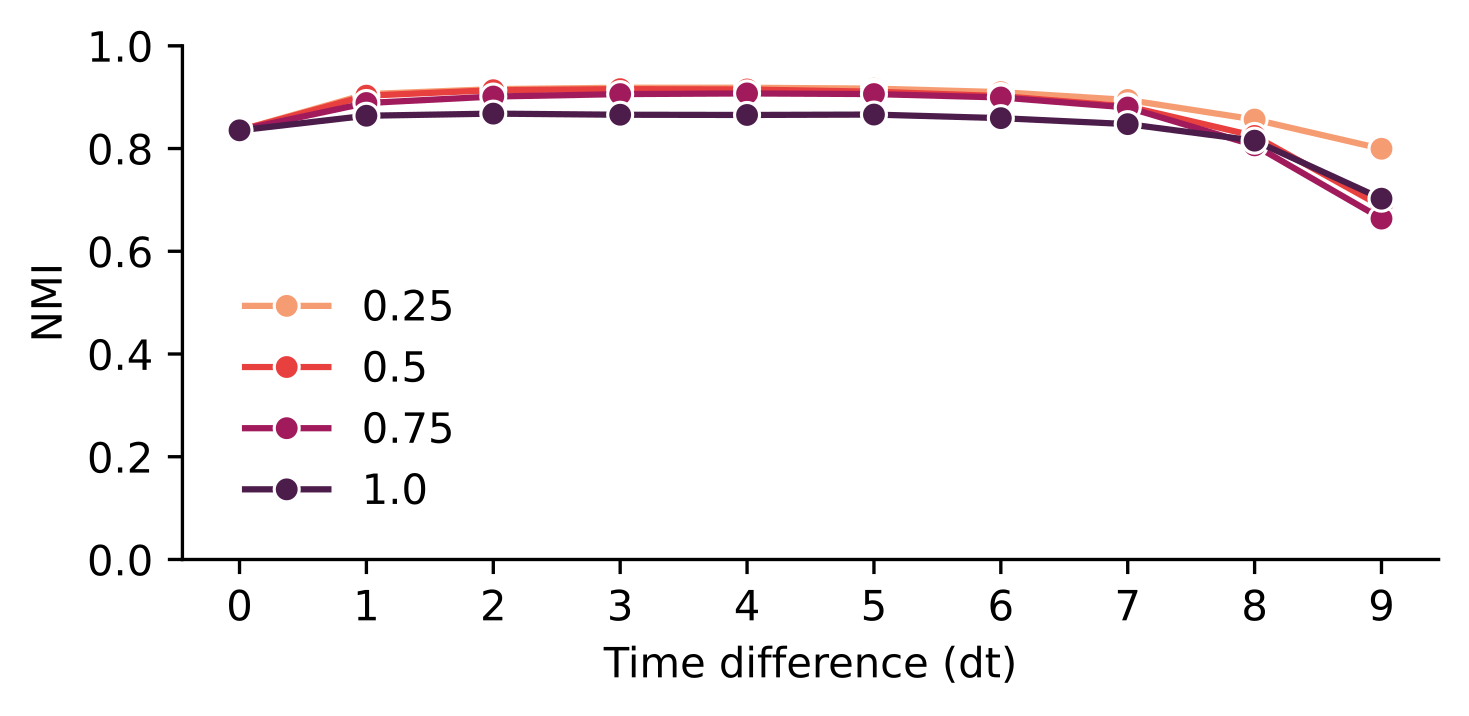}
\caption{NMI vs. time difference for various core node ratios (Louvain algorithm)}
\label{fig:core-nodes-impact}
\end{subfigure}
\hfill
\begin{subfigure}[b]{0.48\textwidth}
\centering
\includegraphics[width=\textwidth]{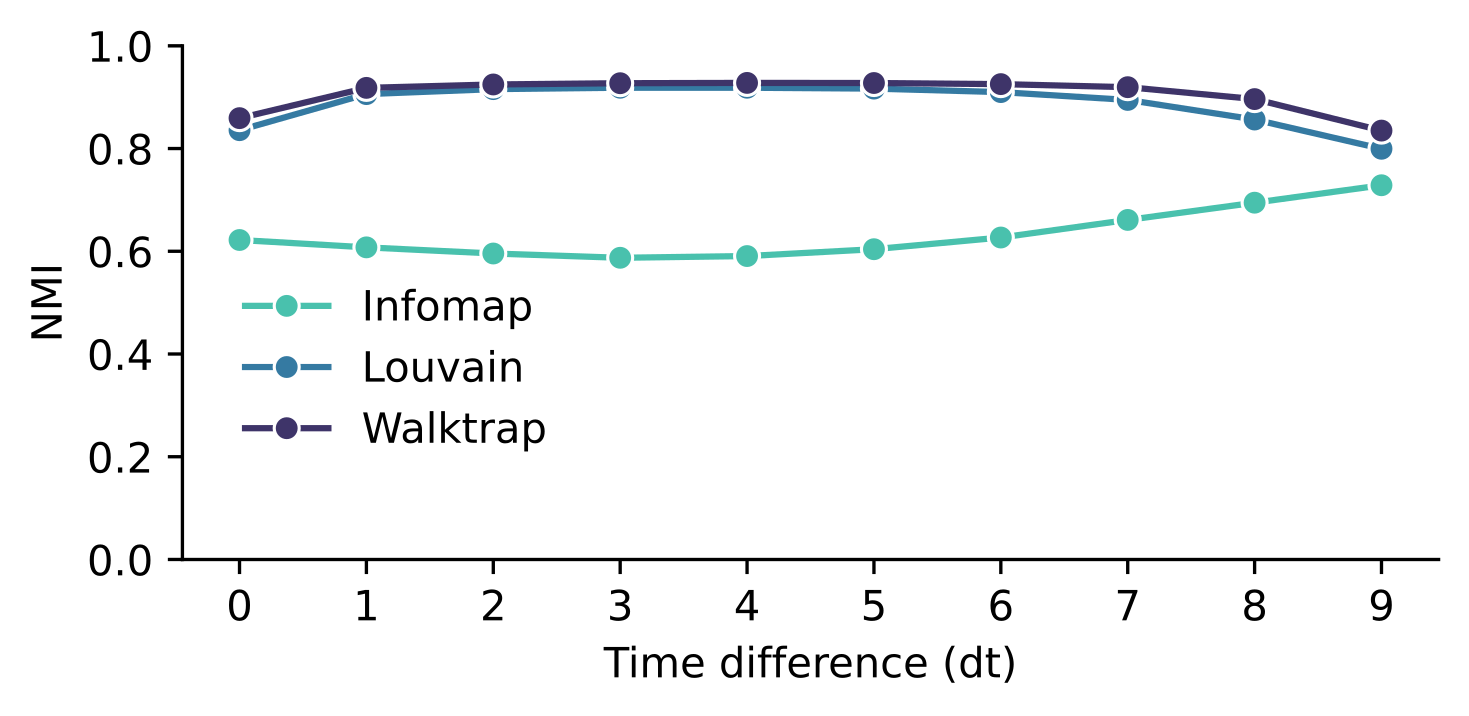}
\caption{NMI vs. time difference for the 3 algorithms (core node ratio = 0.25)}
\label{fig:algo-comparison}
\end{subfigure}
\caption{Temporal tracking performance}
\label{fig:combined-analysis}
\end{figure}

Fig. \ref{fig:combined-analysis}\subref{fig:core-nodes-impact} demonstrates the sensitivity of Louvain to variations in the core nodes ratio. The results show that the algorithm maintains high NMI values (above 0.8) for different core nodes ratios (0.25 to 1.0). 
Fig. \ref{fig:combined-analysis}\subref{fig:algo-comparison} presents a comparative analysis of Louvain, Infomap, and Walktrap with a core nodes ratio of 0.25. The results reveal different patterns: while Louvain and Walktrap demonstrate similar performance with consistently high NMI values (approximately 0.9), Infomap shows notably lower performance (NMI around 0.6) throughout the temporal range. This performance gap suggests that Infomap may be more sensitive to temporal evolution and lose track of membership, while Louvain and Walktrap better preserve community composition across time steps. 

\subsection{Assessing Events}

The ICEM algorithm \cite{KadkhodaMohammadmosaferi2020} is used to detect twelve types of community events. The analysis focuses on the \textit{changing\_0.5} experiment scenario, comparing the performance of Louvain and Infomap algorithms in detecting these events. Other parameters, such as $p_{in}$ and $p_{out}$, are varied, and the results are averaged over all runs.

\begin{figure*}[h!]
    \centering
    \begin{subfigure}{\linewidth}
        \includegraphics[width=\linewidth]{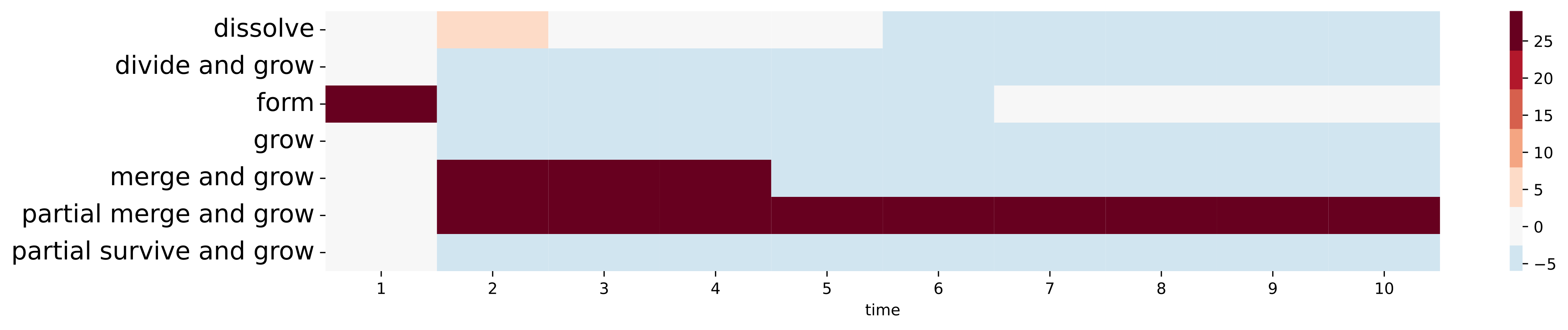}
        \caption{Louvain}
        \label{figure:events:louvain}
    \end{subfigure}
    \begin{subfigure}{\linewidth}
        \centering
        \includegraphics[width=\linewidth]{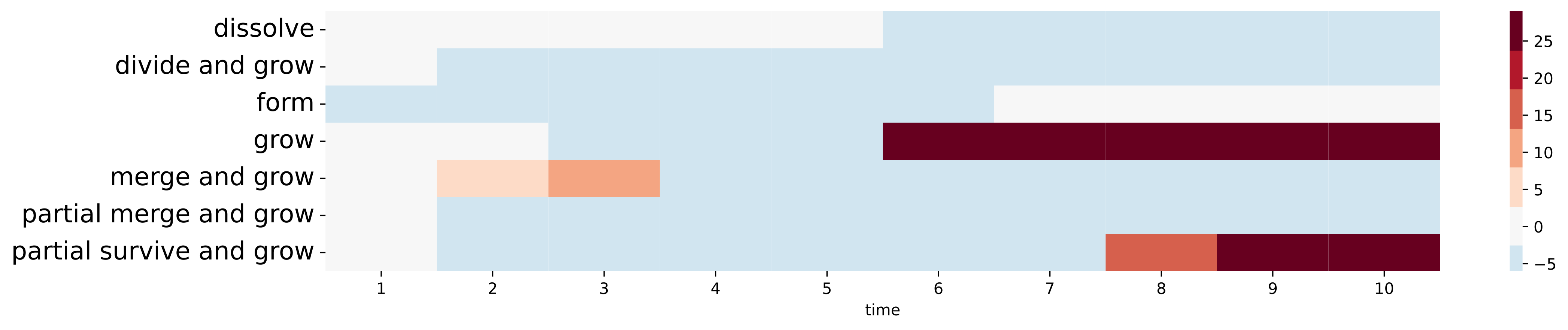}
        \caption{Infomap}
        \label{figure:events:infomap}
    \end{subfigure}
    \caption{Differences between number of events detected by each algorithm and the number of events in the ground truth.}
    \label{figure:events}
\end{figure*}

Fig. \ref{figure:events} shows the differences in event detection between algorithms and the ground truth. The color scale represents detection errors: red indicates an overestimation of events, blue an underestimation, and white areas show accurate detection. The heatmaps reveal distinct error patterns: Louvain overestimates 'form' events early on, 'merge and grow' events in the middle periods, and consistently overestimates 'partial merge and grow' events throughout. Infomap accurately detects 'form' events but overestimates 'merge and grow' events early and 'grow' and 'partial survive and grow' events later.

Both algorithms accurately detect certain events at specific times: Louvain excels with 'dissolve' and 'divide and grow' events, while Infomap performs well with 'dissolve', 'divide and grow', and 'partial merge and grow' events. The choice of algorithm significantly affects the detection of community evolution events, with neither algorithm being universally superior.

\section{Conclusion}

In this paper, we present a novel, highly configurable benchmark for dynamic community detection in temporal networks. Unlike existing models, our framework offers unprecedented flexibility by using probability distributions to generate realistic evolving communities and the corresponding artificial networks that align with these communities and their interactions. Its configurability enables researchers to systematically test algorithms against precisely defined ground truth scenarios with diverse evolutionary behaviors, filling a significant gap in the current literature.

Our benchmark benefits the scientific community in several ways. It enables standardized evaluation of new temporal community detection approaches.
It serves as an experimental platform for studying the impact of network properties on algorithm performance and helps researchers make informed algorithm selections based on specific application needs.
By evaluating community detection algorithms across three dimensions---partition quality, transition tracking, and event detection---our benchmark reveals nuanced algorithmic behaviors. It highlights the importance of considering multiple evaluation criteria, as algorithms may show unexpected strengths in certain aspects despite apparent limitations in others. The entire framework, including data generation, evaluation metrics, and visualization tools, is available as an open-source package \cite{gitlab}.

Future work will focus on incorporating additional graph generation models with scale-free properties, exploring more diverse dynamic events based on empirical studies of real-world evolving communities, and extending the evaluation framework to handle overlapping communities and continuous-time temporal networks, such as link streams.

\bibliographystyle{spmpsci}
\bibliography{refs}

\end{document}